\documentstyle[12pt]{article}
\def\be{\begin{eqnarray}}
\def\ee{\end{eqnarray}}
\def\ba{\begin{array}}
\def\ea{\end{array}}
\def\p{\phi}

\def\a{\alpha}
\def\lam{\lambda}
\def\Lam{\Lambda}
\def\laI{\lambda_{1j}}
\def\laII{\lambda_{2j}}

\def\epII{\epsilon_2}
\def\epI{\epsilon_1}
\def\pa{\partial}

\def\G{{\cal G}}
\def\B{{\cal B}}
\def\A{{\cal A}}
\def\X{{\cal X}}

\def\D{^{(D)}}

\begin{document}
\begin{center}
{\LARGE
{Charging a Double Kerr Solution in 5D \\
\vskip 0.3cm
Einstein--Maxwell--Kalb--Ramond Theory }}
\end{center}

\vskip 1.5cm

\begin{center}
{\bf \large {Ricardo Becerril,}$^{\natural,}$} \footnote{E--mail
address: becerril@ifm.umich.mx}\\
\end{center}
\begin{center} and
\end{center}
\begin{center}
{\bf \large {Alfredo
Herrera--Aguilar}$^{\natural,}$$^{\ddagger,}$} \footnote{E--mail
address: aherrera@auth.gr}
\end{center}
\begin{center}
{$^{\natural}$\it
Instituto de F\'\i sica y Matem\'aticas\\
Universidad Michoacana de San Nicol\'as de Hidalgo\\
Edificio C--3, Ciudad Universitaria, Morelia, Mich., CP 58040
M\'exico}\\
\end{center}
\begin{center}
{$^\ddagger$\it
Theoretical Physics Department, Aristotle University of Thessaloniki\\
54124 Thessaloniki, Greece}\\
\end{center}

\begin{abstract}
We consider the low--energy effective action of the 5D
Einstein--Maxwell--Kalb--Ramond theory. After compactifying this
truncated model on a two--torus and switching off the $U(1)$
vector fields of this theory, we recall a formulation of the
resulting three--dimensional action as a double Ernst system
coupled to gravity. Further, by applying the so--called normalized
Harrison transformation on a generic solution of this double Ernst
system we recover the $U(1)$ vector field sector of the theory.
Afterward, we compute the field content of the generated charged
configuration for the special case when the starting Ernst
potentials correspond to a pair of interacting Kerr black holes,
obtaining in this way an exact field configuration of the 5D
Einstein--Maxwell--Kalb--Ramond theory endowed with effective
Coulomb and dipole terms with momenta. Some physical properties of
this object are analyzed as well as the effect of the normalized
Harrison transformation on the double Kerr seed solution.
\end{abstract}

\newpage
\section{Introduction}
Recently some natural interest has been shown to the study of
field configurations that describe interacting black holes coupled
to some matter fields, both in the framework of General Relativity
\cite{chms}--\cite{vch} and string theory \cite{hk2}--\cite{aha1}.
One of the reasons for such an interest is the development reached
in the statistical approach to physics of single black holes (for
a review, see for instance, \cite{youm}--\cite{peet}) and its
possible generalization to more complicated systems of interacting
black holes coupled to matter.

In this paper we construct a charged field configuration that
consists of a pair of interacting sources of black hole type
coupled to an anti--symmetric Kalb--Ramond tensor field and a set
of Abelian gauge fields in the framework of the truncated
five--dimensional Einstein--Maxwell--Kalb--Ramond (EMKR) theory.
The construction is carried out by applying the normalized
Harrison charging symmetry, which acts on the target space of the
effective three--dimensional heterotic string theory and preserves
the asymptotic properties of the starting field configurations, on
a seed solution that corresponds to a double Ernst system in the
framework of the toroidally reduced five--dimensional
Einstein--Kalb--Ramond (EKR) theory. Several interesting results
have been achieved regarding the physical properties of
five--dimensional black objects \cite{5d}; it turns out that the
BPS bound of rotating black holes is saturated precisely in five
or more dimensions.

The paper is organized as follows: in Section 2 we briefly review
the matrix Ernst potential (MEP) formalism for the effective field
theory of the heterotic string (for an arbitrary number of
dimensions) and its formal analogy to the stationary
Einstein--Maxwell (EM) system. It turns out that after setting to
zero the dilaton and all $U(1)$ vector fields, and considering the
compactification of this theory on a two--torus, the resulting
three--dimensional subsystem admits a K\"ahler representation
which is defined by two vacuum Ernst potentials.

In Section 3 the parametrization which gives rise to this double
Ernst system is pointed out and a discrete transformation between
the metric and Kalb--Ramond degrees of freedom is established. In
Section 4 we recall the normalized Harrison transformation (NHT)
and apply it on a generic seed solution of the EKR theory which
corresponds to two complex Ernst potentials in order to get a
charged field configuration and recover, in this way, the $U(1)$
vector field sector of the EMKR theory.

Further, in Section 5 we reduce the system to two effective
dimensions (dependence of just two dynamical coordinates) in order
to be able to consider as seed solution a pair of Ernst potentials
which correspond to interacting Kerr black holes. In this case,
the $5$--dimensional line element explicitly depends on the Ernst
potentials and the resulting field configuration contains a
Kalb--Ramond dipole hidden inside a horizon. In Section 6 we
explicitly compute the generated charged solution, study its
asymptotical behaviour and give an interpretation of the field
configuration. Finally, we sketch our conclusions and discuss on
the further development of the present work.

\section{Matrix Ernst Potential Formalism}
In this Section we review the MEP formalism for the
$D$--dimensional effective field theory of the heterotic string
and indicate an algorithm for generating a charged solution of the
double Ernst system starting from a neutral one by making use of a
matrix Lie--B\"acklund transformation of Harrison type.

We consider the effective action of the heterotic string theory at
tree level \be S\D\!=\!\int\!d\D\!x\!\mid\!
G\D\!\mid^{\frac{1}{2}}\!e^{-\p\D}\!(R\D\!+\!
\p\D_{;M}\!\p^{(D);M} \!-\!\frac{1}{12}\!H\D_{MNP}H^{(D)MNP}\!-\!
\frac{1}{4}F^{(D)I}_{MN}\!F^{(D)IMN}), \label{2.1} \ee where \be
F^{(D)I}_{MN}\!=\!\pa_MA^{(D)I}_N\!-\!\pa _NA^{(D)I}_M, \quad
H\D_{MNP}\!=\!\pa_MB\D_{NP}\!-\!\frac{1}{2}A^{(D)I}_M\,F^{(D)I}_{NP}\!+\!
\mbox{{\rm \, cycl \, perms \,\, of} \,\, M,\,N,\,P.} \nonumber
\ee Here $G\D_{MN}$ is the metric, $B\D_{MN}$ is the
anti--symmetric Kalb-Ramond field, $\p\D$ is the dilaton,
$A^{(D)I}_M$ is a set of $U(1)$ vector fields ($I=1,\,2,\,...,n$),
$D$ is the original number of space--time dimensions; capital
letters $M,N,...,P$ are related to the whole set of space--time
coordinates, lowercase letters $m,n$ label the extra dimensions,
whereas Greek letters $\mu,\nu$ stand for the non--compactified
coordinates. In the consistent critical case $D=10$ and $n=16$,
but we shall leave these parameters arbitrary for the time being
and will fix them later in Section 3. In \cite{marsch}--\cite{ms}
it was shown that after the compactification of this model on a
$D-3=d$--torus, the resulting three--dimensional theory possesses
the $SO(d+1,d+n+1)$ symmetry group that later was identified as
$U$--duality \cite{u} and describes gravity through the metric
tensor \be g_{\mu\nu}\!=\!e^{-2\p}\!\left(G\D_{\mu\nu}\!-
\!G\D_{m+3,\mu}G\D_{n+3,\nu}G^{mn}\right),\nonumber\ee coupled to
the following set of three--dimensional fields:

\noindent a) scalar fields \be G\!\equiv\!G_{mn}\!=
\!G\D_{m+3,n+3},\,\,\, B\!\equiv\!B_{mn}\!= \!B\D_{m+3,n+3},\,\,\,
A\!\equiv\!A^I_m\!= \!A^{(D)I}_{m+3},\,\,\,
\p\!=\!\p\D\!-\!\frac{1}{2}{\rm ln|det}\,G|, \ee \noindent
b)tensor field \be
B_{\mu\nu}\!=\!B\D_{\mu\nu}\!\!-\!4B_{mn}A^m_{\mu}A^n_{\nu}\!-\!
2\!\left(A^m_{\mu}A^{m+d}_{\nu}\!-\!A^m_{\nu}A^{m+d}_{\mu}\right),
\ee \noindent c)vector fields $A^{(a)}_{\mu}=
\left((A_1)^m_{\mu},(A_2)^{m+d}_{\mu},(A_3)^{2d+I}_{\mu}\right)$
\be (A_1)^m_{\mu}\!=\!\frac{1}{2}G^{mn}G\D_{n+3,\mu},\,
(A_3)^{I+2d}_{\mu}\!=\!-\frac{1}{2}A^{(D)I}_{\mu}\!+\!A^I_nA^n_{\mu},\,
(A_2)^{m+d}_{\mu}\!=\!\frac{1}{2}B\D_{m+3,\mu}\!\!-\!B_{mn}A^n_{\mu}\!+\!
\frac{1}{2}A^I_{m}A^{I+2d}_{\mu} \ee where the subscripts
$m,n=1,2,...,d$; and $a=1,...,2d+n$. In this letter we set
$B_{\mu\nu}=0$ since an anti--symmetric tensor has no dynamical
degrees of freedom in three dimensions; this is equivalent to
removing the effective cosmological constant which, in general, is
included in the spectrum of the three--dimensional effective
theory.

We dualize all vector fields on--shell with the aid of the
pseudoscalar fields $u$, $v$ and $s$ as follows:
\begin{eqnarray}
\nabla\times\overrightarrow{A_1}&=&\frac{1}{2}e^{2\p}G^{-1}
\left(\nabla u+(B+\frac{1}{2}AA^T)\nabla v+A\nabla s\right),
\nonumber                          \\
\nabla\times\overrightarrow{A_3}&=&\frac{1}{2}e^{2\p}
(\nabla s+A^T\nabla v)+A^T\nabla\times\overrightarrow{A_1},
\label{2.5}\\
\nabla\times\overrightarrow{A_2}&=&\frac{1}{2}e^{2\p}G\nabla v-
(B+\frac{1}{2}AA^T)\nabla\times\overrightarrow{A_1}+
A\nabla\times\overrightarrow{A_3}.
\nonumber
\end{eqnarray}
Thus, the effective three--dimensional theory describes gravity
$g_{\mu\nu}$ coupled to the scalars $G$, $B$, $A$, $\p$ and
pseudoscalars $u$, $v$, $s$. In \cite{hk3} it was shown that all
these matter fields can be arranged in the following pair of MEP
\be \X= \left( \ba{cc}
-e^{-2\p}+v^TXv+v^TAs+\frac{1}{2}s^Ts&v^TX-u^T \cr Xv+u+As&X \ea
\right), \quad \qquad \A=\left( \ba{c} s^T+v^TA \cr A \ea \right),
\label{2.6} \ee where $X=G+B+\frac{1}{2}AA^T$, in such a way that
they reproduce the field equations of the three--dimensional
theory. These matrices have dimensions $(d+1) \times (d+1)$ and
$(d+1) \times n$, respectively.

In terms of the MEP the effective three--dimensional theory adopts
the form \be ^3S\!= \!\int\!d^3x\!\mid
g\mid^{\frac{1}{2}}\!\{\!-\!R\!+ \!{\rm
Tr}[\frac{1}{4}\left(\nabla \X\!-\!\nabla \A\A^T\right)\!\G^{-1}
\!\left(\nabla \X^T\!-\!\A\nabla \A^T\right)\!\G^{-1}
\!+\!\frac{1}{2}\nabla \A^T\G^{-1}\nabla \A]\}, \label{2.7} \ee
where $\X=\G+\B+\frac{1}{2}\A\A^T$, then \,
$\G=\frac{1}{2}\left(\X+\X^T-\A\A^T\right)$ and \be \G= \left(
\ba{cc} -e^{-2\p}+v^TGv&v^TG \cr Gv&G \ea \right), \quad \B=\left(
\ba{cc} 0&v^TB-u^T \cr Bv+u&B \ea \right). \label{2.8} \ee In
\cite{hk3} it also was shown that there exist a map between the
stationary actions of the heterotic string and EM theories. The
map reads \be \X\longleftrightarrow -E, \quad
\A\longleftrightarrow F, \nonumber \ee \be {\it matrix\,\,
transposition}\quad\longleftrightarrow\quad {\it complex\,\,
conjugation}, \label{2.9} \ee where $E$ and $F$ are the
conventional complex Ernst potentials of the stationary EM theory
\cite{e1}. This map allows us to extrapolate the results obtained
in the EM theory to the heterotic string realm using the MEP
formulation.
\subsection{The normalized Harrison transformation}
In the language of the MEP the three--dimensional action (7)
possesses a set of symmetries which has been classified according
to their charging properties in \cite{hk5}. Among them one finds
the matrix Ehlers and Harrison transformations \cite{ehl}, which
are symmetries that change the properties of the spacetime in a
non--trivial way; they represent the matrix counterpart of the
B\"acklund transformation of the sine--Gordon equation in the
realm of the stationary heterotic string theory. For instance, the
so--called normalized Harrison transformation allows us to
construct charged string vacua from neutral ones preserving the
asymptotical values of the three--dimensional seed fields. Namely,
the matrix transformation \be
&&\A\rightarrow\left(1+\frac{1}{2}\Sigma\lambda\lambda^T\right)
\left(1-\A_0\lambda^T+\frac{1}{2}\X_0\lambda\lambda^T\right)^{-1}
\left(A_0-\X_0\lambda\right)+\Sigma\lambda, \label{2.10}
\\
&&\X\rightarrow\left(1+\frac{1}{2}\Sigma\lambda\lambda^T\right)
\left(1-\A_0\lambda^T+\frac{1}{2}\X_0\lambda\lambda^T\right)^{-1}
\left[\X_0+\left(\A_0-\frac{1}{2}\X_0\lambda\right)\lambda^T\Sigma\right]
+\frac{1}{2}\Sigma\lambda\lambda^T\Sigma, \nonumber \ee where
$\Sigma={\rm diag}(-1,-1,1,...,1)$ stands for the signature that
the MEP $\X$ adopts at spatial infinity and $\lambda$ is an
arbitrary constant $(d+1)\times n$--matrix, generates charged
string solutions (with non--zero potential $\A$) from neutral ones
if we start from the seed potentials \be \X_0\ne 0, \qquad \qquad
\A_0=0. \nonumber \ee The parameters that enter the matrix
$\lambda$ can be interpreted as electromagnetic charges that
couple to the original seed object. It is precisely with the aid
of this B\"acklund transformation that we shall charge the double
5D Ernst system in the next Section.

\section{5D Einstein-Kalb-Ramond vs double Ernst system}
In this Section we present a formulation of the resulting
three--dimensional model, upon toroidal compactification of the 5D
EKR theory, as a double Ernst system by means of a complete
parametrization of the matrices $\G$ and $\B$ in terms of the real
and imaginary parts of a pair of complex Ernst potentials.

Let us begin by setting to zero all the $U(1)$ gauge fields which
correspond to the winding modes of the three--dimensional theory
(this is equivalent to dropping the matrix $\A$ in (7)). Thus, we
obtain the following action in terms of the MEP $\X$ \be ^3S\!=
\!\int\!d^3x\!\mid g\mid^{\frac{1}{2}}\!\left\{\!-\!R\!+
\frac{1}{4}\!{\rm Tr}\,\left[ \nabla \X\!\G^{-1}\!\nabla
\X^T\!\G^{-1}\right]\right\} =\!\int\!d^3x\!\mid
g\mid^{\frac{1}{2}}\!\left\{\!-\!R\!+ \frac{1}{4}\!{\rm
Tr}\,\left(J^{\X}J^{\X^{T}}\right)\right\} \label{3.1} \ee where
now $\X=\G+\B$,\,\, $\G=\frac{1}{2}\left(\X+\X^T\right)$ and
$J^{\X}=\nabla \X\G^{-1}$.

There are two physically different effective theories that can be
expressed by the action (\ref{3.1}), and hence admit a double
Ernst formulation. On the one hand we have the $D=5$ EKR model,
where the dilaton field is set to zero as well \cite{hk2}. On the
other hand we have the $D=4$ bosonic string theory, for which a
charged pair of rotating interacting black holes coupled to
dilaton and Kalb--Ramond fields was constructed in \cite{aha1} and
its charged dual string vacua were studied in \cite{hn}. Here we
will consider again the 5D EKR theory in order to apply the NHT on
a neutral family of field configurations that correspond to the
double Ernst system.

Thus, we start with the five--dimensional truncated action
\be
^5S=\int d\,^5x\mid^5\!\!G\mid^{\frac{1}{2}}\left(^5\!R-
\frac{1}{12}^5\!\!H^2 \right),
\label{3.2}
\ee
where $^5R$ is the Ricci scalar constructed on the
$5$--dimensional metric $^5G_{MN}$ and
\begin{equation}
^5H_{MNP}={\partial_{M}}^5\!\!B_{NP}
+ {\rm cyc.\,\, perms. \,\,of \,\,M,N,P.}
\end{equation}
It is worth noticing that we are considering a truncation which imposes the
following condition on the Kaluza--Klein and Kalb--Ramond vector fields
\begin{equation}
^5G_{\mu,n+2}=^5\!\!B_{\mu,n+2}=0;
\label{3.4}
\end{equation}
this implies that the vector fields $A_1$ and $A_2$ must vanish
identically, and hence, the pseudoscalar fields $u$ and $v$ also
vanish (see (\ref{2.5})). Such a
restriction does not provide any constraint on the remaining dynamical
variables and can be considered as a consistent non--trivial ansatz for
the EKR theory.

After the Kaluza--Klein reduction on $T^2$ we get the stationary
effective action (\ref{3.1}) (see, for instance, \cite{hk2},
\cite{ms}) with the matter field spectrum of the theory encoded in
the $(2\times 2)$--matrices $\G\equiv G$ and $\B\equiv B$ which
can be parametrized in the following form \be \G= \frac{p_1}{p_2}
\left( \ba{cc} 1&q_2\cr q_2&p_2^2+q_2^2 \ea \right), \qquad \qquad
\B= q_1 \left(
\begin{array}{crc}
0  & \quad & -1\\
1 & \quad & 0\\
\end{array}
\right)=q_1\sigma_2,
\label{3.5}
\ee
where $\sigma_2$ is the Pauli matrix. Under such assumptions,
the five--dimensional interval reads
\begin{equation}
ds_5^2 =g_{\mu\nu}dx^{\mu}dx^{\nu}+\G_{mn}dx^mdx^n.
\label{3.6}
\end{equation}

By substituting (\ref{3.5}) into (\ref{3.1}) the action of the
``matter fields" adopts the form
\begin{eqnarray}
^3 S_m = \frac{1}{2}
\int d^3x {\mid g \mid}^{\frac {1}{2}} \left\{
p_1^{-2}\left[(\nabla p_1)^2 + (\nabla q_1)^2\right] +
p_2^{-2}\left[(\nabla p_2)^2 + (\nabla q_2)^2\right]
\right\},
\label{3.7}
\end{eqnarray}
which allows us to introduce two independent Ernst potentials
\begin{equation}
\epsilon _1 =p_1+iq_1, \qquad \epsilon _2=p_2+iq_2.
\label{3.8}
\end{equation}
In terms of these field variables, the action of the system can be rewritten as
a double Ernst system in the K\"ahler form \cite{mazur}:
\begin{eqnarray}
^3 S =
\int d^3x {\mid g \mid}^{\frac {1}{2}} \left\{ - ^3 R +
2\left(J^{\epsilon _1}J^{\overline\epsilon _1} +
J^{\epsilon _2}J^{\overline\epsilon _2} \right) \right\},
\label{3.9}
\end{eqnarray}
where
$J^{\epsilon _1}=\nabla\epsilon _1\,(\epsilon _1+\overline\epsilon _1)^{-1}$
and
$J^{\epsilon _2}=\nabla\epsilon _2\,(\epsilon _2+\overline\epsilon _2)^{-1}$.

A mathematically equivalent, but physically different $2 \times 2$--matrix
representation arises from (12) by making use of the discrete symmetry
$p_1 \longleftrightarrow p_2$, \, $q_1 \longleftrightarrow q_2$.
This fact allows us to define new matrices
\begin{eqnarray}
\G'=
\frac{p_2}{p_1}
\left (\begin{array}{crc}
1& \quad & q_1\\
q_1 & \quad & p_1^2+q_1^2\\
\end{array} \right),
\qquad \qquad
\B'=
q_2\sigma_2
\label{3.10}
\end{eqnarray}
and, hence, $\X'=\G'+\B'$ and to write down the action that corresponds to
these magnitudes:
\be
^3S\!=
\!\int\!d^3x\!\mid g\mid^{\frac{1}{2}}\!\left\{\!-\!R\!+
\frac{1}{4}\!{\rm Tr}\,\left(J^{\X'}J^{\X'^{T}}\right)\right\}
\!=\!\int\!d^3x\!\mid g\mid^{\frac{1}{2}}\!\left\{\!-\!R\!+
2\left(J^{\epsilon '_1}J^{\overline\epsilon '_1}+
J^{\epsilon '_2}J^{\overline\epsilon '_2}\right)\right\},
\label{3.11}
\ee
where similarly $J^{\X'}=\nabla\X'\G'^{-1}$, \,
$J^{\epsilon '_1}=\nabla\epsilon '_1\,(\epsilon '_1+\overline\epsilon '_1)^{-1}$,\,
$J^{\epsilon '_2}=\nabla\epsilon '_2\,(\epsilon '_2+\overline\epsilon '_2)^{-1}$,\,
$\epsilon '_1=p_2+iq_2$ and $\epsilon '_2=p_1+iq_1$.

In terms of the MEP the above--mentioned discrete transformation
reads \be \X \longleftrightarrow \X'; \label{3.12} \ee thus, the
matrices $\G'$ and $\B'$ must be interpreted as new Kaluza--Klein
and Kalb--Ramond fields, respectively. This symmetry mixes the
gravitational and matter degrees of freedom of the theory. It
recalls the Bonnor transformation of the EM theory \cite{b}, but
in the bosonic string realm. It can be used to generate new
solutions starting, for instance, from pure Kaluza--Klein string
vacua (see \cite{hn} as well).

\section{Applying the NHT on the Double Ernst System}
Let us now proceed to apply the NHT on the neutral double Ernst
system. This will generate a non--zero electromagnetic potential
$\A$ which accounts for non--trivial Abelian $U(1)$ gauge fields.
In order to achieve this aim, we must consider the following seed
MEP \be \X_0= \left( \ba{ccc}
\frac{p_1}{p_2}&\quad&\frac{p_1q_2-q_1p_2}{p_2}\cr
\quad&\quad&\quad\cr \frac{p_1q_2+q_1p_2}{p_2}&\quad&
\frac{p_1}{p_2}(p_2^2+q_2^2) \ea \right), \qquad \qquad \A_0=0.
\label{4.1} \ee In this case, the charge matrix $\lam$ that
parametrizes the NHT has the form \be \lam= \left( \ba{cccc}
\lam_{11}&\lam_{12}&...&\lam_{1n}\cr
\lam_{21}&\lam_{22}&...&\lam_{2n}\cr \ea \right), \label{4.2} \ee
where $n\ge 2$ for consistency. Thus after applying the NHT on
this double Ernst seed solution, the transformed MEP read \be
\X_{11}\!=\!\frac{1}{\Xi}\left[\left(4+\Lam^2 |\epII |^2\right)
Re\epI+2\left(\laI^2+\laII^2 |\epI |^2\right)Re\epII + 4\laI\laII
Re\epII Im\epI\right], \label{4.3} \ee \be
\X_{12}\!=\!\frac{1}{\Xi}\left\{\Gamma_{+}\left( Re\epI
Im\epII-Re\epII Im\epI\right)+2\laI\laII\left[(1-|\epI
|^2)Re\epII-(1-|\epII |^2)Re\epI\right]\right\}, \ee \be
\X_{21}\!=\!\frac{1}{\Xi}\left\{\Gamma_{-} \left(Re\epI
Im\epII+Re\epII Im\epI\right)+2\laI\laII\left[(1-|\epII
|^2)Re\epI+(1-|\epI |^2)Re\epII\right]\right\}, \ee \be
\X_{22}\!=\!\frac{1}{\Xi}\left[ \left(\Lam^2 +4|\epII |^2\right)
Re\epI+2\left(\laII^2+\laI^2 |\epI |^2\right)Re\epII - 4\laI\laII
Re\epII Im\epI\right], \ee \be \A_{1j}\!=\!\frac{2}{\Xi}\left\{
\left[\left(2-\laII^2 |\epI |^2\right)Re\epII-\left(2-\laII^2
|\epII|^2\right)Re\epI + \laI\laII\left(Re\epI Im\epII-Re\epII
Im\epI\right)\right]\lam_{1j}-\right. \nonumber \ee \be
\left.\left[\left(2+\laI^2\right)\left(Re\epI Im\epII-Re\epII
Im\epI\right)+\laI\laII\left(|\epII |^2Re\epI- |\epI
|^2Re\epII\right)\right]\lam_{2j} \right\}, \label{4.4} \ee \be
\A_{2j}\!=\!\frac{-2}{\Xi}\left\{\left[\left(2+\laII^2\right)
\left(Re\epI Im\epII+Re\epII Im\epI\right)+\laI\laII \left(Re\epI
-|\epI |^2 Re\epII\right)\right]\lam_{1j}\right.- \nonumber \ee
\be \left.\left[\left(\laI^2-2|\epII |^2\right)Re\epI +
\left(2-\laI^2 |\epI |^2\right)Re\epII + \laI\laII\left(Re\epI
Im\epII+Re\epII Im\epI\right)\right]\lam_{2j}\right\}, \label{4.5}
\ee \be \Xi=2\left(\laI^2+\laII^2 |\epII |^2\right)Re\epI +
\left(4+\Lam^2 |\epI |^2\right)Re\epII + 4\laI\laII Re\epI
Im\epII, \ee where
$\Lam^2=\laI^2\lambda_{2j}^2-\left(\laI\laII\right)^2$,
$\Gamma_{+}=4+2\laI^2-2\laII^2-\Lam^2$,
$\Gamma_{-}=4-2\laI^2+2\laII^2-\Lam^2$ and the non--trivial
character of the matrix $\A$ is evident. The fields configurations
corresponding to these potentials live now in the 5D EMKR theory
since we have recovered the $U(1)$ vector fields of the system.

\section{Double Kerr seed solution}
In this section, following \cite{hk2} we impose one more symmetry
on the fields of the three--dimensional effective theory under
consideration in order to use as seed solution a pair of Kerr
black holes. Thus, we can write the line element in the
Lewis--Papapetrou form making use of the Weyl coordinates as
follows \be ^5\!ds^2=\G_{mn}dx^mdx^n+
e^{2\gamma}\left(d\rho^2+dz^2\right)-\rho^2d\tau^2 \label{5.1} \ee
where $\G_{mn}$ and $\gamma$ are $\tau$--independent. Thus, a
solution of our system can be constructed using the solutions of
the double vacuum Einstein equations written in the Ernst form in
terms of $\epsilon _k$ and $\gamma ^{\epsilon _k}$ ($k=1,2$) \be
\nabla (\rho J^{\epsilon _k})&=& \rho J^{\epsilon _k}(J^{\epsilon
_k}-J^{\bar \epsilon _k}),
\nonumber\\
\partial _{z} \gamma ^{\epsilon _k}&=&
\rho \left [(J^{\epsilon _k})_z (J^{\bar \epsilon _k})_{\rho}
+(J^{\bar \epsilon _k})_z (J^{\epsilon _k})_{\rho}\right ],
\label{5.2}\\
\partial _{\rho} \gamma ^{\epsilon _k}&=&
\rho \left [{|(J^{\epsilon _k})_{\rho}}|^2
-{|(J^{\epsilon _k})_z}|^2\right],
\nonumber
\ee
if one identifies the function $\gamma$ that accounts for the general
relativistic interaction between de black holes, in the following way
$\gamma \equiv \gamma ^{\epsilon _1} + \gamma ^{\epsilon _2}$.

For instance, we can take as seed solution a double Kerr system
consisting of a pair of rotating interacting black holes. In the
framework of General Relativity, the Ernst potentials
corresponding to two Kerr solutions with sources in different
points of the symmetry axis read:
\begin{eqnarray}
\epsilon_k = 1 - \frac {2m_k}{r_k + i\a_k\cos\theta_k}.
\label{5.3}
\end{eqnarray}
where $m_k$ and $\a_k$ are constant parameters which define the masses and
rotations of the sources of the Kerr field configurations.
Weyl and Boyer--Lindquist coordinates are related through
\begin{eqnarray}
\rho = [(r_k - m_k)^2-\zeta _k^2]^{\frac{1}{2}}\sin\theta_k,
\qquad z = z_k + (r_k - m_k)\cos\theta_k, \label{5.4}
\end{eqnarray}
where the sources are located at $z_k$ and
$\zeta _k^2 = m_k^2 - a_k^2$. Thus, for the function
$\gamma _k$ we have
\begin{equation}
e^{2\gamma _k} = \frac {P_k}{Q_k}, \label{5.5}
\end{equation}
where $P_k = \Delta_k - \a_k^2 \sin^2 \theta _k$, \, $Q_k = \Delta_k +
\zeta _k^2 \sin^2 \theta _k$
and $\Delta _k = r_k^2 - 2m_k r_k + \a_k^2$.

We would like to make a remark at this point: when parameterizing
the 5D interval (\ref{5.1}) in the Lewis--Papapetrou form, one
could choose a completely spatial (Euclidean) three--dimensional
interval and require that the signature of the matrix $\G$ to be
negative definite, i.e., $\G|_{\infty}=-I_2$ (the same signature
holds for the matrix $\X|_{\infty}$). Thus, the five--dimensional
metric would possess a signature with two time--like coordinates.
Such kind of models have been studied in \cite{bars} and represent
another line of investigation within this approach. It is clear
that in order to fulfill this condition either $\epsilon_1$ or
$\epsilon_2$ must adopt the asymptotic value $-1$, since when both
potentials have the same asymptotic behaviour (with the same sign)
the signature of the matrix $\G$ is positive definite.

Thus, in the language of the complex Ernst potentials the field
configuration adopt the form \be ds_5^2=
e^{2\gamma}(d\rho^2+dz^2)-\rho^2d\tau^2+
\frac{\epsilon_1+\bar\epsilon_1}{\epsilon_2+\bar\epsilon_2}
\left|du+i\bar\epsilon_2dv\right|^2, \label{5.6} \ee
\begin{eqnarray}
\B=\frac{\epsilon_1-\bar\epsilon_1}{2i}\sigma_2, \label{5.7}
\end{eqnarray}
where $u=x^4$, $v=x^5$ and
$\gamma=\gamma^{\epsilon_1}+\gamma^{\epsilon_2}$ as it was pointed
out above.

In the case when the Ernst potentials correspond to two
interacting Kerr black holes, the symmetric matrix $\G$ is
determined by the following relations \be \G_{uu}=
\frac{(r_1^2-2m_1r_1+\a_1^2\cos^2\theta_1)(r_2^2+\a_2^2\cos^2\theta_2)}
{(r_2^2-2m_2r_2+\a_2^2\cos^2\theta_2)(r_1^2+\a_1^2\cos^2\theta_1)},
\nonumber \ee \be \G_{uv}=
\frac{2m_2\a_2\cos\theta_2(r_1^2-2m_1r_1+\a_1^2\cos^2\theta_1)}
{(r_2^2-2m_2r_2+\a_2^2\cos^2\theta_2)(r_1^2+\a_1^2\cos^2\theta_1)},
\label{5.8} \ee \be \G_{vv}=
\frac{(r_1^2-2m_1r_1+\a_1^2\cos^2\theta_1)
(r_2^2-4m_2r_2+4m_2^2+\a_2^2\cos^2\theta_2)}
{(r_2^2-2m_2r_2+\a_2^2\cos^2\theta_2)(r_1^2+\a_1^2\cos^2\theta_1)},
\nonumber \ee the factor $e^{2\gamma}$ reads \be e^{2\gamma}
=\frac{(r_1^2-2m_1r_1+\a_1^2\cos^2\theta_1)
(r_2^2-2m_2r_2+\a_2^2\cos^2\theta_2)}
{(r_1^2-2m_1r_1+\a_1^2\cos^2\theta_1+m_1^2\sin^2\theta_2)
(r_2^2-2m_2r_2+\a_2^2\cos^2\theta_2+m_2^2\sin^2\theta_2)},
\label{5.9} \ee and the Kalb--Ramond matrix $\B$ is defined as \be
\B=\frac{2m_1\a_1\cos\theta_1}{r_1^2+\a_1^2\cos^2\theta_1}\sigma_2
\label{5.10} \ee and can be interpreted as a matrix Kalb--Ramond
dipole configuration with momentum $m_1\a_1$ located at $z_1$ and
hidden inside the horizon $r_1=m_1+\sqrt{m_1^2-\a_1^2}$ of the
metric (\ref{5.6}). Simultaneously, the $\G_{uv}$ metric component
also constitutes a dipole configuration but possesses momentum
$m_2\a_2$ and is located at $z_2$, hidden inside the horizon
$r_2=m_2+\sqrt{m_2^2-\a_2^2}$.

\section{Charged Field Configurations in 5D EMKR Theory}
After applying the NHT on the double Ernst seed solution we get
the following field configurations: \be
\G_{uu}=\X_{11}-\frac{1}{2}\A_{1j}^2, \qquad
\G_{uv}=\frac{1}{2}\left(\X_{12}+\X_{21}-\A_{1j}\A_{2j}\right),
\qquad \G_{vv}=\X_{22}-\frac{1}{2}\A_{2j}^2, \label{6.1} \ee \be
\B=\frac{1}{2}\left(\X_{21}-\X_{12}\right)\sigma_2,  \qquad
\A\equiv A= \left( \ba{c} \A_{1j}\cr \A_{2j} \ea \right),
\label{6.2} \ee where the appearance of the electromagnetic
potential is obvious. By substituting the Ernst potentials
$\epsilon_k$ by the corresponding double Kerr black hole system we
obtain the following charged field configuration \be \G_{uu}=
\frac
{DQ\Delta_1\Delta_2\!+\!4m_1\!\left(\!L^2r_1\!+\!2m_1\laII^2\!+\!
2\laI\laII\a_1\cos\theta_1\!\right)\!\Delta_2\!
+\!2m_2\!\left[(4\!-\!\Lambda^2)r_2\!+\!2\Lambda^2m_2\right]\!\Delta_1}
{DQ\Delta_1\Delta_2\!+\!4m_2\!\left(\!L^2r_2\!+\!2m_2\laII^2\!+\!
2\laI\laII\a_2\cos\theta_2\!\right)\!\Delta_1\!+\!
2m_1\!\left[(4\!-\!\Lambda^2)r_1\!+\!2\Lambda^2m_1\right]\!\Delta_2}-
\nonumber \ee \be \frac {8\left(h_1\laI-h_2\laII\right)^2+
2\Lambda^2\left(h_2\laI-h_3\laII\right)^2-
8\Lambda^2\left(h_1h_3-h_2^2\right)}
{\left[DQ\Delta_1\Delta_2\!+\!4m_2\!\left(\!L^2r_2\!+\!2m_2\laII^2\!+\!
2\laI\laII\a_2\cos\theta_2\!\right)\!\Delta_1\!+\!
2m_1\!\left[(4\!-\!\Lambda^2)r_1\!+\!2\Lambda^2m_1\right]\!\Delta_2
\right]^2}, \label{6.3} \ee \be \G_{uv}= \frac
{4m_1\left[2\laI\laII(r_1-m_1)-
L^2\a_1\cos\theta_1\right]\Delta_2+
2(4-\Lambda^2)m_2\a_2\cos\theta_2\Delta_1}
{DQ\Delta_1\Delta_2\!+\!4m_2\!\left(\!L^2r_2\!+\!2m_2\laII^2\!+\!
2\laI\laII\a_2\cos\theta_2\!\right)\!\Delta_1\!+\!
2m_1\!\left[(4\!-\!\Lambda^2)r_1\!+\!2\Lambda^2m_1\right]\!\Delta_2}+
\nonumber \ee \be
\frac{\!4\!\left(\!l_1^2h_1h_4\!+\!l_2^2h_2h_6\!\right)\!
+\!2\Lam^2\!\left[\!\left(\!2\!+\!\laI^2\!\right)\!h_2h_5\!
-\!\left(\!2\!+\!\laII^2\!\right)\!h_3h_4\right]\!
-\!2\laI\laII\!\left[4h_1h_6\!+\!\Lam^2h_3h_5\!+\!\left(\!4-\!\Lam^2\!\right)\!
h_2h_4\right]} {\left[DQ\Delta_1\Delta_2\!+\!
4m_2\!\left(\!L^2r_2\!+\!2m_2\laII^2\!+\!
2\laI\laII\a_2\cos\theta_2\!\right)\!\Delta_1\!+\!
2m_1\!\left[(4\!-\!\Lambda^2)r_1\!+\!2\Lambda^2m_1\right]\!\Delta_2\right]^2},
\label{6.4} \ee \be \G_{vv}= \frac {DQ\Delta_1\Delta_2\!
-\!2m_2\!\left[(4\!-\!\Lambda^2)r_2\!-\!8m_2\right]\!\Delta_1\!
-\!4m_1\!\left(\!L^2r_1\!-\!2m_1\laI^2\!
+\!2\laI\laII\a_1\cos\theta_1\!\right)\!\Delta_2}
{DQ\Delta_1\Delta_2\!+\!4m_2\!\left(\!L^2r_2\!+\!2m_2\laII^2\!+\!
2\laI\laII\a_2\cos\theta_2\!\right)\!\Delta_1\!+\!
2m_1\!\left[(4\!-\!\Lambda^2)r_1\!+\!2\Lambda^2m_1\right]\!\Delta_2}-
\nonumber \ee \be \frac
{\!8\!\left(\!h_4\!\laI\!-\!h_6\laII\!\right)^2\!+\!
2\Lambda^2\left(h_5\laI+h_4\laII\!\right)^2\!+
\!8\Lambda^2\left(h_5h_6+h_4^2\right)}
{\left[DQ\Delta_1\Delta_2\!+\!4m_2\!\left(\!L^2r_2\!+\!2m_2\laII^2\!+\!
2\laI\laII\a_2\cos\theta_2\!\right)\!\Delta_1\!+\!
2m_1\!\left[(4\!-\!\Lambda^2)r_1\!+\!2\Lambda^2m_1\right]\!\Delta_2
\right]^2}, \label{6.5} \ee \be \B_{uv}\!=\!
\frac{2\left(4-\Lambda^2\right)m_1\a_1\cos\theta_1
\Delta_2+4m_2\left[2\laI\laII(r_2-m_2)-
L^2\a_2\cos\theta_2\right]\Delta_1}
{DQ\Delta_1\Delta_2\!+\!4m_2\!\left(\!L^2r_2\!+\!2m_2\laII^2\!+\!
2\laI\laII\a_2\cos\theta_2\!\right)\!\Delta_1\!+\!
2m_1\!\left[(4\!-\!\Lambda^2)r_1\!+\!2\Lambda^2m_1\right]\!\Delta_2},
\label{6.6} \ee \be \A_{1j}= \frac
{2\left[2h_1-\lam_{2i}^2h_3+\lam_{1i}\lam_{2i}h_2\right]\laI+
2\left[\lam_{1i}\lam_{2i}h_3-\left(2+\lam_{1i}^2\right)h_2\right]\laII}
{DQ\Delta_1\Delta_2\!+\!4m_2\!\left( \!L^2r_2\!+\!2m_2\laII^2\!+\!
2\laI\laII\a_2\cos\theta_2\!\right)\!\Delta_1\!+\!
2m_1\!\left[(4\!-\!\Lambda^2)r_1\!+\!2\Lambda^2m_1\right]\!\Delta_2},
\label{6.7} \ee \be \A_{2j}= \frac
{-2\left[\left(2+\lam_{2i}^2\right)h_4+\lam_{1i}\lam_{2i}h_5\right]\laI+
2\left[\lam_{1i}^2h_5+\lam_{1i}\lam_{2i}h_4+2h_6\right]\laII}
{DQ\Delta_1\Delta_2\!+\!4m_2\!\left( \!L^2r_2\!+\!2m_2\laII^2\!+\!
2\laI\laII\a_2\cos\theta_2\!\right)\!\Delta_1\!+\!
2m_1\!\left[(4\!-\!\Lambda^2)r_1\!+\!2\Lambda^2m_1\right]\!\Delta_2},
\label{6.8} \ee where \be
h_1=2\left(m_1r_1\Delta_2-m_2r_2\Delta_1\right), \qquad
h_2=2\left(m_2\a_2\cos\theta_2\Delta_1-m_1\a_1\cos\theta_1\Delta_2\right),
\nonumber \ee \be
h_3=4\left(m_1^2\Delta_2-m_2^2\Delta_1\right)-h_1, \qquad
h_4=2\left(m_1\a_1\cos\theta_1\Delta_2+m_2\a_2\cos\theta_2\Delta_1\right),
\label{6.9} \ee \be
h_5=2\left[m_1(r_1-2m_1)\Delta_2+m_2r_2\Delta_1\right], \qquad
h_6=2\left[m_1r_1\Delta_2+m_2(r_2-2m_2)\Delta_1\right], \nonumber
\ee $L^2=\laI^2-\laII^2$, \,\,$l_k^2=2\lam_{kj}^2+\Lam^2$,
\,\,$\Delta_k=r_k^2-2m_kr_k+\a_k^2\cos^2\theta_k$, $(k=1,2)$ and,
finally, $DQ=4+2\laI^2+2\laII^2+\Lambda^2$.

A consistency checking of the generated solution consists of
setting the parameters $\laI$ and $\laII$ to zero in order to
recover the starting field configuration
(\ref{5.8})--(\ref{5.10}). It is straightforward to verify that
this is indeed the case.

The asymptotical behaviour of the generated three--dimensional
field configurations read \be \label{6.10} \G_{uu}|_{\infty}\sim
1-\frac{2\Gamma_-(m_1-m_2)}{DQr}+\frac{8\laI\laII
\left(m_1\a_1\cos\theta_1-m_2\a_2\cos\theta_2\right)}{DQr^2}+{\cal
O}(r^{-2}), \ee \be \label{6.11}
\G_{uv}|_{\infty}\sim\frac{8\laI\laII m_1}{DQr}-
\frac{4L^2m_1\a_1\cos\theta_1-
2\left(4-\Lam^2\right)m_2\a_2\cos\theta_2}{DQr^2}+{\cal
O}(r^{-2}), \ee \be \label{6.12}\G_{vv}|_{\infty}\sim
1-\frac{2\Gamma_+(m_1+m_2)}{DQr}-\frac{8\laI\laII
\left(m_1\a_1\cos\theta_1+m_2\a_2\cos\theta_2\right)}{DQr^2}+{\cal
O}(r^{-2}), \ee \be \B_{uv}|_{\infty}\sim\frac{8\laI\laII
m_2}{DQr}+\frac{2\left(4-\Lam^2\right)m_1\a_1\cos\theta_1-
4L^2m_2\a_2\cos\theta_2}{DQr^2}+{\cal O}(r^{-2}), \label{6.13} \ee
\be \A_{1j}|_{\infty}\sim
\frac{4\left[(2+\lam_{2i}^2)\laI-\lam_{1i}\lam_{2i}\laII\right](m_1-m_2)}
{DQr}- \nonumber\ee \be
\frac{4\left[\lam_{1i}\lam_{2i}\laI-(2+\lam_{1i}^2)\laII\right]
\left(m_1\a_1\cos\theta_1-m_2\a_2\cos\theta_2\right)}{DQr^2}+{\cal
O}(r^{-2}),\label{6.14} \ee \be
\A_{2j}|_{\infty}\sim\frac{-4\left[\lam_{1i}\lam_{2i}\laI-
(2+\lam_{1i}^2)\laII\right](m_1+m_2)}{DQr}+ \nonumber\ee \be
\frac{4\left[(2+\lam_{2i}^2)\laI-\lam_{1i}\lam_{2i}\laII\right]
\left(m_1\a_1\cos\theta_1+m_2\a_2\cos\theta_2\right)}{DQr^2}+{\cal
O}(r^{-2}). \label{6.15} \ee From this analysis it is clear that
under the NHT, all the generated fields (gravitational,
Kalb--Ramond and electromagnetic) effectively develop both Coulomb
and dipole terms. Thus, from one side, the $\G_{uu}$ component of
the constructed metric possesses mass terms defined by
$M_{uu_1}=\Gamma_{-}m_1/DQ$ at $z_1$ and
$M_{uu_2}=\Gamma_{-}m_2/DQ$ at $z_2$, and from the other side, it
acquires dipole sources with masses $\widetilde
M_{uu_1}=8\laI\laII m_1/DQ$, $\widetilde M_{uu_2}=-8\laI\laII
m_2/DQ$, and their corresponding momenta $\widetilde
M_{uu_1}\a_1$, $\widetilde M_{uu_2}\a_2$, located at $z_1$ and
$z_2$, respectively. In a similar way the $\G_{vv}$ component of
the metric has masses $M_{vv_1}=\Gamma_{+}m_1/DQ$ at $z_1$ and
$M_{vv_2}=\Gamma_{+}m_2/DQ$ at $z_2$; indeed, it possesses as well
the massive dipole terms defined by $\widetilde
M_{vv_1}=-8\laI\laII m_1/DQ$ at $z_1$ and $\widetilde
M_{uu_2}=-8\laI\laII m_2/DQ$ at $z_2$ with the momenta $\widetilde
M_{vv_1}\a_1$ and $\widetilde M_{vv_2}\a_2$, respectively.

The transformed Kalb--Ramond tensor field also acquires a Coulomb
term determined by the charge $M_B=8\laI\laII m_2/DQ$ located at
$z_2$ and now possesses two dipole sources with masses
$M_{B_1}=\left(4-\Lam^2\right)m_1/DQ$, $M_{B_2}=-4L^2m_2/DQ$ and
momenta $M_{B_1}\a_1$, $M_{B_2}\a_2$, located at $z_1$ and $z_2$,
respectively. The same situation exactly takes place for the
$\G_{uv}$ component of the metric, which usually corresponds to
the rotation of the gravitational field. Thus, the generated
gravitational potential $\G_{uv}$ has a Coulomb source with mass
$M_{uv}=8\laI\laII m_1/DQ$ located at $z_1$ and dipole sources
with masses $\widetilde M_{uv_1}=-2L^2m_1/DQ$, $\widetilde
M_{uv_2}=\left(4-\Lam^2\right)m_2/DQ$ and momenta $\widetilde
M_{uv_1}\a_1$, $\widetilde M_{uv_2}\a_2$, located at $z_1$ and
$z_2$, respectively.

At this point we would like to point out that the discrete
symmetry (\ref{3.12}) which relates gravitational and Kalb--Ramond
degrees of freedom is still present asymptotically and is quite
evident in the language of the masses and charges of the
components $\G_{uv}$ and $\B_{uv}$, since one can clearly see that
these components transform into each other under the interchange
of the respective masses and charges, even after the
implementation of the nonlinear NHT.

Finally, the generated field configuration possesses an evidently
non--trivial electromagnetic sector and its asymptotic structure
reveals its usual Coulomb form, defining in this way the effective
electromagnetic charges of the system. These fields also have
effective dipole sources. Thus, the electromagnetic fields
$\A_{1j}$ possess momenta defined by the expressions
$4\left[\lam_{1i}\lam_{2i}\laI-(2+\lam_{1i}^2)\laII\right]m_1\a_1/DQ$
and
$-4\left[\lam_{1i}\lam_{2i}\laI-(2+\lam_{1i}^2)\laII\right]m_2\a_2/DQ$,
whereas the respective momenta for the electromagnetic fields
$\A_{2j}$ read
$4\left[(2+\lam_{2i}^2)\laI-\lam_{1i}\lam_{2i}\laII\right]m_1\a_1/DQ$
and
$4\left[(2+\lam_{2i}^2)\laI-\lam_{1i}\lam_{2i}\laII\right]m_2\a_2/DQ$.

Thus, under the NHT, the double Kerr seed solution does not
acquire just the electromagnetic charges, but it develops as well
effective Coulomb and dipole terms for all the fields of the field
configuration: gravitational, Kalb--Ramond and electromagnetic
fields.

\section{Conclusion and Discussion}
In this paper we have obtained a charged field configuration of
the five--dimensional EMKR theory starting from a neutral one that
corresponds to a double Ernst (double Kerr, in particular) system.
The generation of the new charged solution was carried out via a
matrix Lie--B\"acklund transformation of Harrison type that
preserves the asymptotical values of the seed fields.

An interesting novel feature of the generated exact solution is
that all the fields of the field configuration develop effective
Coulomb and dipole terms asymptotically. Thus, after applying the
NHT, the 5D double Kerr seed solution acquires effective Coulomb
terms and dipole sources with momenta. This is in contrast with
the effect that the NHT produces on a neutral seed solution in the
framework of the general theory of relativity where it just endows
the initial field configuration with a set of electromagnetic
charges.

The statistical analysis of such a configuration is an appealing
direction to conduct the present research. The equilibrium
properties of the generated solution is of interest as well and
would generalize to the 5D case some previous results obtained in
the framework of the four--dimensional general relativity
\cite{vch}, \cite{tk}--\cite{mrsg}. These issues are under current
investigation.

\section{Acknowledgements}
One of the authors (AHA) is really grateful to the Theoretical
Physics Department of the Aristotle University of Thessaloniki
and, specially, to Prof. J.E. Paschalis for useful discussions and
for providing a stimulating atmosphere while part of this work was
being done. AHA is also really grateful to S. Kousidou for her
support while this research was in progress and acknowledges a
grant for postdoctoral studies provided by the Greek Government.
Both authors thank IMATE--UNAM campus Morelia and CINVESTAV--IPN
for library facilities provided during this investigation. RB's
work was supported by grant CIC-UMSNH-4.11, while AHA's research
by grants CONACYT-J34245-E, CONACYT-42064-F and CIC-UMSNH-4.16.

\end{document}